\documentclass[aps,pra,showpacs,twocolumn]{revtex4-1}
\usepackage{amssymb}
\usepackage{amsmath}
\usepackage{graphicx}
\usepackage{epsfig}

\begin{document}

\title{Dynamic manifestation of exception points in a non-Hermitian
continuous model with an imaginary periodic potential}
\author{Y. T. Wang}
\author{R. Wang}
\email{wangr@tjnu.deu.cn}
\author{X. Z. Zhang}
\email{zhangxz@tjnu.edu.cn}
\affiliation{College of Physics and Materials Science, Tianjin Normal
University,Tianjin 300387, China}

\begin{abstract}
Exceptional points (EPs) are distinct characteristics of non-Hermitian
Hamiltonians that have no counterparts in Hermitian systems. In this study,
we focus on EPs in continuous systems rather than discrete non-Hermitian
systems, which are commonly investigated in both the experimental and
theoretical studies. The non-Hermiticity of the system stems from the local
imaginary potential, which can be effectively achieved through particle loss
in recent quantum simulation setups. Leveraging the discrete Fourier
transform, the dynamics of EPs within the low-energy sector can be well
modeled by a Stark ladder system under the influence of a non-Hermitian
tilted potential. To illustrate this, we systematically investigate
continuous systems with finite imaginary potential wells and demonstrate the
distinctive EP dynamics across different orders. Our investigation sheds
light on EP behaviors, potentially catalyzing further exploration of EP
phenomena across a variety of quantum simulation setups.
\end{abstract}

\maketitle

\affiliation{College of Physics and Materials Science, Tianjin Normal
University,Tianjin 300387, China}

\section{Introduction}

The establishment of quantum mechanics is based on the Hermiticity of the
Hamiltonian operator, which ensures the reality of the eigenvalues and the
orthogonality of the eigenvectors. However, this mathematical framework is
typically used to analyze closed systems that conserve energy and
probability. In 1998, the pioneering work \cite{Bender1} by Bender
demonstrated that a class of non-Hermitian Hamiltonians with parity-time
reversal symmetry can possess a fully real spectrum. Subsequently, Ali \cite%
{Ali} illustrated that pseudo-Hermitian operators are crucial for realizing
the reality of the spectrum. Recently, non-Hermitian systems have been
simulated in various physical systems, spanning from atomic \cite%
{Bloch,LLH1,LQ,LLH2} to photonic \cite{FL,PM,OSK,ZXY,LAD} platforms. Despite
their ability to exhibit real spectra, these systems are not closed,
allowing for energy and particle exchange with the environment. Disruption
of the balance between the system and its environment leads to the emergence
of a complex spectrum. Within this process, a critical region must exist
where typical dynamic behavior is altered, accompanied by the appearance of
an EP \cite{KT,HWD1}. Furthermore, the development of non-Hermitian
topological phases \cite{BEJ} has introduced a new dimension to traditional
condensed matter physics and quantum optics, reshaping our understanding of
matter \cite{BEJ,LS1,EGR,MMA,PM2,AY,ZSM,WH,FC}.

Recent advancements in experimental dissipation manipulation have reignited
interest in studying open quantum systems, where dissipative processes are
fundamental for quantum state preparation \cite{VF,MM,BJT}. Specifically,
experimental employment of dissipative coupling can achieve Tonks-Girardeau
gas of molecules \cite{SN} and topological states \cite{DS}, probing
peculiar dynamical behaviors tied to passive parity-time symmetry in
dissipative quantum systems \cite{LJ,NM,DL,LX}. The ability to simulate
non-Hermitian systems in open quantum setups has led to a series of
discoveries, including non-Hermitian skin effects \cite{LTE,YS,KFK,AVMM,LCH}
and topological classifications exceeding the standard ten classes \cite%
{GZ,KK,ZH,LCH1,LCH2}. Recent studies have unveiled distinct dynamic traits
of non-Hermitian skin effects in dissipative systems \cite%
{LLH1,LQ,ZXY,Longhi,LCH3}, alongside the appearance of chiral damping \cite%
{SF}, helical damping \cite{LCH4}, and edge bursts \cite{XWT}. The motion of
electrons in a lattice under an electric field is often elucidated by the
Stark ladder model, involving a linear potential. This model is feasible in
a tilted optical lattices \cite{DMB,WSR}, showcasing intriguing phenomena
like Bloch oscillations \cite{BF,Longhi2}, Zener tunneling \cite{ZC,Longhi3}%
, and many-body Stark localization \cite{SM,MW}. Recent research has
explored the Stark ladder model influenced by dissipative imaginary linear
potentials, broadening the horizons of conventional Stark ladder model
investigations \cite{CS}.

When addressing problems in continuous quantum systems, it is necessary to
discretize the system to a moderate extent to facilitate computation by
computers. The Fourier grid Hamiitonian (FGH) method maps the kinetic and
potential energy components of the Hamiltonian onto grid points in
coordinate space. Through Fourier transformations, it ultimately yields a
sum of cosine functions for each matrix element of the Hamiltonian in
coordinate space \cite{M1989}. It has proved a remarkably easy and robust
method for computing the vibrational motion of one-dimensional systems \cite%
{KB1992,PRL2018}. Systems processed with the FGH method more closely
approximate real physical systems and thus have significant reference value
for experiments. However, the application of the FGH method to the treatment
of non-Hermitian quantum systems remains unexplored.

In this study, we expand the FGH method to encompass non-Hermitian
continuous systems influenced by tilted imaginary potentials. Within the
low-energy sector, dynamics are effectively modulated by a non-Hermitian
tight-binding system, enabling the identification of EPs. We observe
distinct orders of EPs within continuous systems featuring various wells.
Notably, a two-order EP, termed a scale-free EP, remains unchanged with
increasing system size. The different EP orders exhibit unique dynamic
behaviors discernible through the growth of the total Dirac probability,
following a power law linked to the EP order index. Our findings offer
valuable insights into the EP dynamics of non-Hermitian continuous systems.

Our paper is organized as follows. In Sec. \ref{Discretization of a
non-Hermitian continuum model with complex potential}, we expand the FGH
method to non-Hermitian continuous systems. In Sec. \ref{Effective
Hamiltonian and coalescing states}, we describe the coalescing states of
effective tight-binding Hamiltonian by the discretization of continuous
system. In Sec. \ref{Scale-free EP}, we demonstrate the existence of
scale-free behavior of EPs in the effective Hamiltonian. In Sec. \ref{The
dynamics at EP}, we discuss the different dynamical behaviors of EPs,
including EP2 and EP3. A summary is given in Sec. \ref{Summary}.

\section{Discretization of a non-Hermitian continuum model with complex
potential}

\begin{figure*}[tbp]
\includegraphics[bb=49 80 1506 578,width=1.0\textwidth]{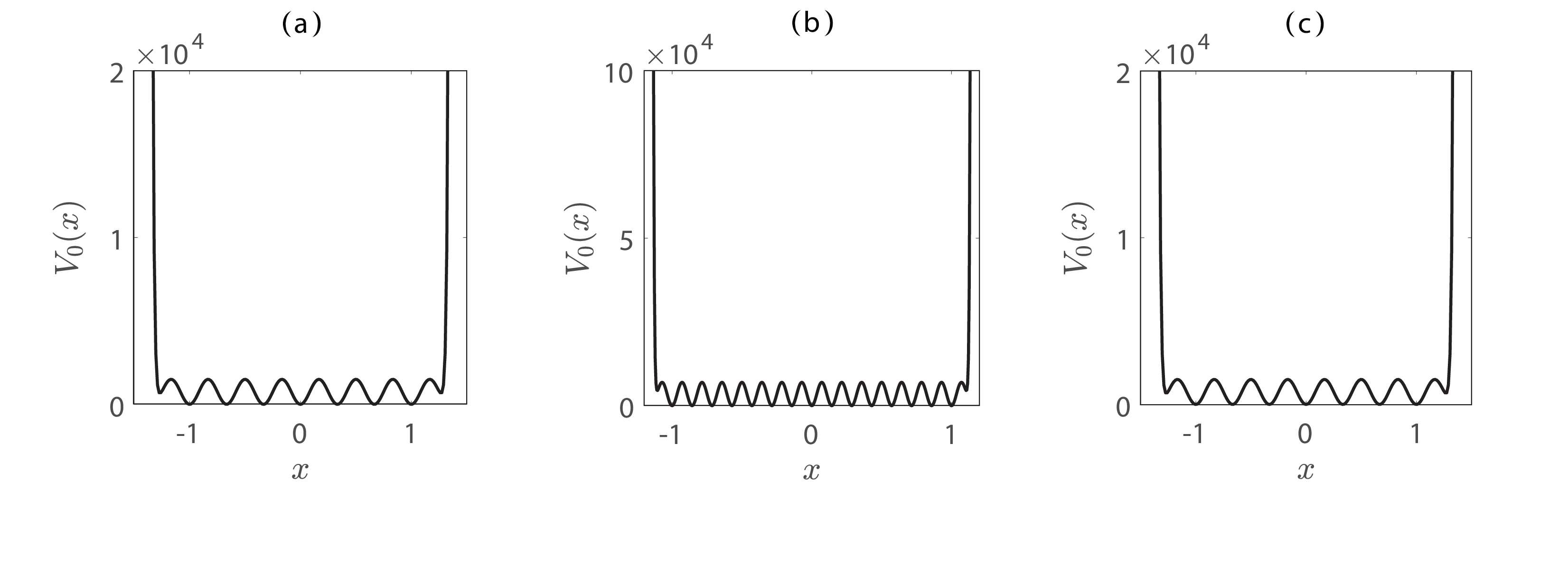}
\caption{(Color online) The continuum system comprises $N$ potential wells
created by $V_{0}\left( x\right) $ with the imaginary potential omitted from
this representation. Figs. \protect\ref{fig1}(a)-(c) correspond to the
system with $N=5,$ $7$, and $15$ potential wells, respectively. The specific
parameter values for each system are as follows: (a) $N^{\prime }=201,$ $L=3,
$ $\protect\omega =2, $ $\protect\gamma =800,$ $b=0.76,$ $a=80$; (b) $%
N^{\prime }=201,$ $L=3,$ $\protect\omega =3,$ $\protect\gamma =1500,$ $%
b=0.83,$ $a=100$; (c) $N^{\prime }=401,$ $L=2.3,$ $\protect\omega =7,$ $%
\protect\gamma =7000,$ $b=0.935,$ $a=200$.}
\label{fig1}
\end{figure*}

\label{Discretization of a non-Hermitian continuum model with complex
potential} First, we examine a single particle exposed to a complex
potential, with the corresponding Hamiltonian expressed as
\begin{equation}
H=T+V\left( x\right) ,
\end{equation}%
where $T=p^{2}/2m$ denotes the kinetic energy, and $V\left( x\right)
=V_{0}\left( x\right) -i\kappa x$ defines the complex potential. Here,
\begin{equation}
V_{0}\left( x\right) =\gamma \sin ^{2}\left( \omega \pi x\right) +\left(
bx\right) ^{a}
\end{equation}%
where $\gamma $ is the depth of the potential well, and $i\kappa x$ is a
linear imaginary potential which can be realized through the lossy system. A
higher value of $\gamma $ results in a deeper potential well. The parameter $%
\omega $ controls the width of each potential well, determining both the
width and the number of wells within a system of a specified length. The
second term sets the boundaries for the potential, confining the well to the
range of $x\in \lbrack -1,1]$. This condition requires that $b\leq 1$, with
the specific value dependent on the width and depth of each potential well.
Notably, if the depth of each well increases and their width decreases, $b$
tends towards $1$ to ensure that the wells at the edge positions have the
same width as those at other positions. For simplicity, $a$ is chosen as a
large, even number. Beyond the interval, i.e., $x\notin \lbrack -1,1]$, this
component tends to diverge, indicating an open boundary for the system.
Further cases are illustrated in Fig. \ref{fig1}, which will be employed to
show the subsequent interesting dynamics.

To discretize the system under consideration, we begin by expressing the
Hamiltonian in coordinate representation, where the matrix element is
defined as
\begin{eqnarray}
\left\langle x|H|x^{\prime }\right\rangle &=&\frac{1}{2\pi }\int_{-\infty
}^{+\infty }\exp [ik(x-x^{\prime })]T_{k}dk  \notag \\
&&+V\left( x\right) \delta \left( x-x^{\prime }\right) ,
\label{H_continuous}
\end{eqnarray}%
where $T_{k}=\hbar ^{2}k^{2}/2m$. Initially, we substitute the discretized
value $x_{n}=n\Delta x$ for the continuous variable $x$, where $%
n=0,1,2,\cdots ,N^{\prime }-1$, $N^{\prime }$ represents the number of
discrete points. Here, $\Delta x=L/\left( N^{\prime }-1\right) $ denotes the
spacing between adjacent points, with $L$ being the length of the system.
The total length $L$ and the spacing $\Delta x$ determine the size of the
reciprocal lattice and also establish the maximum wavelength value.
Consequently, the expression for the reciprocal lattice in momentum space is
\begin{equation}
\Delta k=2\pi /L=2\pi /\left( N^{\prime }-1\right) \Delta x.
\end{equation}%
For convenience, we designate the center coordinate of the momentum space
grid as $k=0$ , ensuring a symmetric distribution of grid points around $k=0$%
. To maintain this symmetry, we opt for an odd number of discrete grid
points $N^{\prime }$ in the coordinate space. Correspondingly, the
discretization in the momentum space can be given by $k_{l}=l\Delta k$,
where $-\tau \leq l\leq \tau ,\tau =(N^{\prime }-1)/2$. In the
discretization process, the identity operator is rewritten as
\begin{equation}
\hat{I}_{x}=\sum_{n=1}^{N^{\prime }}|x_{n}\rangle \Delta x\langle x_{n}|,
\end{equation}%
which satisfies $\Delta x\left\langle x_{n}|x_{n^{\prime }}\right\rangle
=\delta _{nn^{\prime }}$. By converting the integral into a summation and
substituting $\Delta x$ and $\Delta k$ into the Eq. (\ref{H_continuous}), we
can obtain%
\begin{eqnarray}
H_{nn^{^{\prime }}} &=&\frac{2}{\Delta x}\{\sum_{l=1}^{\tau }\cos \left[
2\pi l\left( n-n^{\prime }\right) /(N^{\prime }-1)\right] T_{l}/(N^{\prime
}-1)  \notag \\
&&+V\left( x_{n}\right) \delta _{nn^{\prime }}\}.  \label{H_dis}
\end{eqnarray}%
where $T_{l}=\left( \hbar l\Delta k\right) ^{2}/(2m)$. It is assumed here
that $H_{nn^{\prime }}$ with respect to $l=0$ point.

Next, we turn to investigate the discretized state of the system. In the
continuous limit, the eigenstates of the Hamiltonian $H$ and its adjoint $%
H^{\dagger }$ in coordinate representation are denoted as $\left\langle
x|\psi \right\rangle =\psi \left( x\right) $ and $\left\langle x|\overline{%
\psi }\right\rangle =\overline{\psi }\left( x\right) $, respectively. The
biorthonormalization condition is expressed as
\begin{equation}
\int_{-\infty }^{\infty }\overline{\psi }^{\ast }\left( x\right) \psi \left(
x\right) dx=1.
\end{equation}%
During the discretization process, this expression transforms into
\begin{equation}
\sum_{n=1}^{N^{\prime }}\overline{\psi }^{\ast }(x_{n})\psi (x_{n})\Delta
x=1.
\end{equation}%
Similarly, the wavefunction in coordinate space is given by $\left\langle
x_{n}|\psi \right\rangle =\psi \left( x_{n}\right) =\psi _{n}$ ($%
\left\langle x_{n}|\overline{\psi }\right\rangle =\overline{\psi }\left(
x_{n}\right) =\overline{\psi }_{n}$). Consequently, we have
\begin{equation}
|\psi \rangle =\hat{I}_{x}\left\vert \psi \right\rangle
=\sum_{n=1}^{N^{\prime }}|x_{n}\rangle \Delta x\psi _{n}
\end{equation}%
and%
\begin{equation}
|\overline{\psi }\rangle =\hat{I}_{x}\left\vert \overline{\psi }%
\right\rangle =\sum_{n=1}^{N^{\prime }}|x_{n}\rangle \Delta x\overline{\psi }%
_{n}.
\end{equation}%
The corresponding eigenenergy can be given as
\begin{equation}
E=\frac{\langle \overline{\psi }|H|\psi \rangle }{\langle \overline{\psi }%
|\psi \rangle }=\frac{\sum_{nn^{\prime }}\overline{\psi }_{n}^{\ast }\Delta
xH_{nn^{\prime }}\Delta x\psi _{n}}{\Delta x\sum_{n}|\overline{\psi }%
_{n}\psi _{n}|}.
\end{equation}%
In this regard, the Eq. (\ref{H_dis}) can be reformulated as
\begin{eqnarray}
H_{nn^{\prime }}^{0} &=&2\sum_{l=1}^{\tau }\cos [2\pi l\left( n-n^{\prime
}\right) /(N^{\prime }-1)]T_{l}/(N^{\prime }-1)  \notag \\
&&+V\left( x_{n}\right) \delta _{nn^{\prime }},  \label{H0}
\end{eqnarray}%
where $T_{l}=2\{\hbar \pi l/[\left( N^{\prime }-1\right) \Delta x]\}^{2}/m$.
Referring to Eq. (\ref{H0}), the expectation value of $H_{nn^{\prime }}^{0}$
is rewritten as
\begin{equation}
E=\frac{\sum_{nn^{\prime }}\overline{\psi }_{n}^{\ast }H_{nn^{\prime
}}^{0}\psi _{n}}{\sum_{n}|\overline{\psi }_{n}\psi _{n}|}.
\end{equation}%
At this stage, the discretization of the continuous system is complete, and
the matrix elements of the discretized Hamiltonian can be solved using Eq. (%
\ref{H0}). When discretizing a continuous system with $N^{\prime }$ grid
points, we transform the Hamiltonian into an $N^{\prime }$ order matrix,
yielding $N^{\prime }$ eigenvalues and $N^{\prime }$ eigenvectors.
Typically, increasing the number of grid points $N^{\prime }$ leads to more
precise results. The choice of $N^{\prime }$ is also influenced by the form
of the potential energy $V_{0}\left( x\right) $. In cases where the
discretization length contains multiple potential wells, a larger $N^{\prime
}$ ensures that the grid is fine enough to capture the details of each well.

\section{Effective Hamiltonian and coalescing states}

\label{Effective Hamiltonian and coalescing states} 
\begin{figure*}[tbp]
\includegraphics[bb=16 42 1875 1056,width=1.0\textwidth]{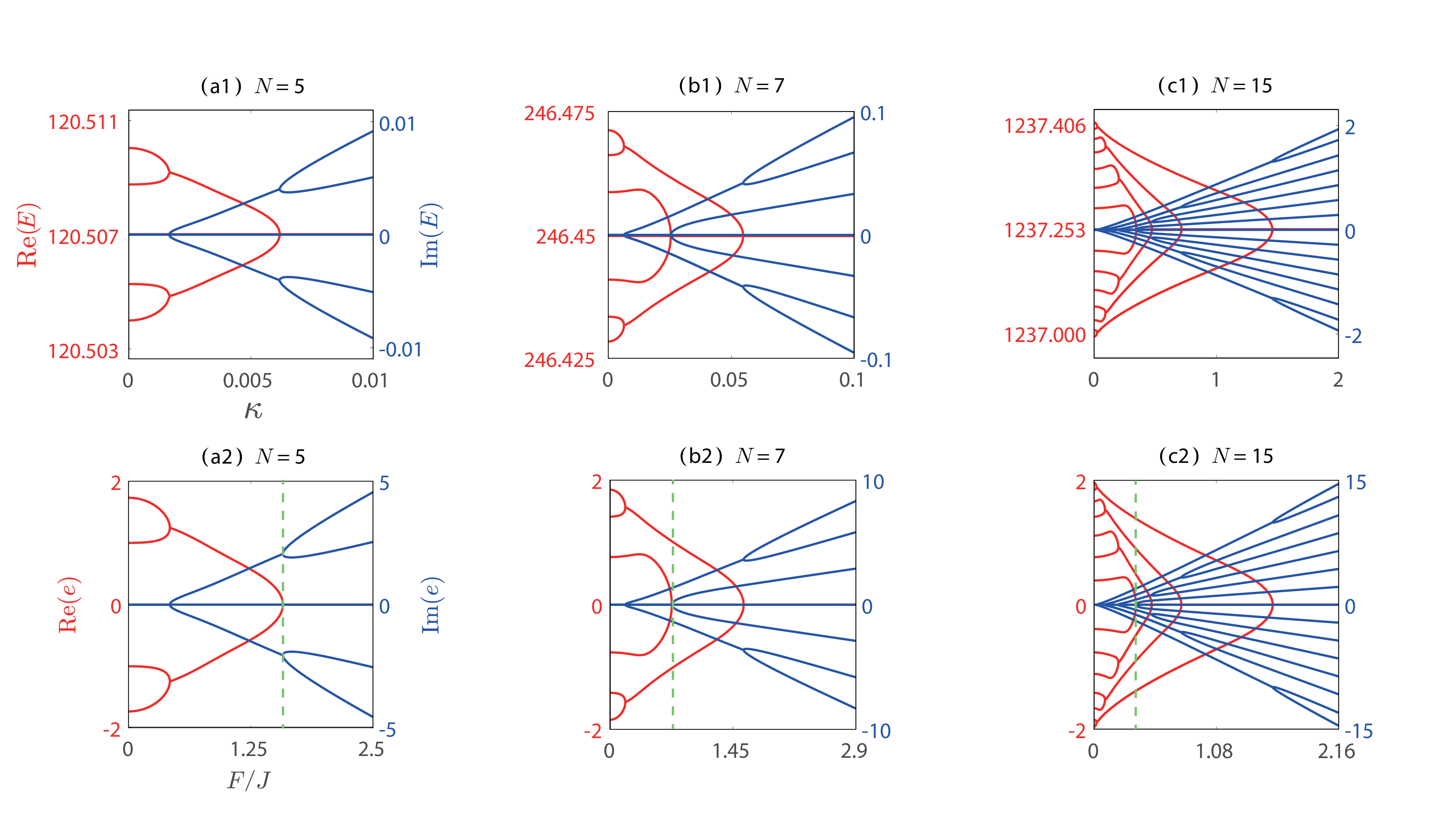}
\caption{(Color online) The low-lying spectrum of the discretized
Hamiltonian $H$ with the potential wells is depicted for $N=5$ in (a1), $N=7$
in (b1), and $N=15$ in (c1), respectively. The red (blue) solid line
represents the real (imaginary) part of the eigenenergy. Figs. \protect\ref%
{fig2}(a2)-(c2) display the eigenspectrum of the effective tight-binding
Hamiltonian $H_{\text{\textrm{eff}}}$. Clearly, the two spectra exhibit the
same structure, demonstrating the accurate depiction of the low-lying
structure of $H$ by the tight-binding Hamiltonian $H_{\text{\textrm{eff}}}$.
Notably, two eigenstates coalesce in (a1), while three eigenstates coalesce
in both Figs. \protect\ref{fig2}(b1)-(c1), corresponding to EP$2$ and EP$3$,
respectively. This low-lying behavior is also evident in Figs. \protect\ref%
{fig2}(a2)-(c2). Here, $J=1$ is assumed for simplicity. }
\label{fig2}
\end{figure*}

In the domain of
periodic potentials, the band structure manifests when particles are exposed
to such potentials. Within the single-band approximation, the system can
exhibit $N$ low-lying eigenenergies that are notably distinct from the
remaining eigenenergies by a discernible gap. This distinguishing feature
persists even in the presence of a complex potential. Our attention is
directed towards this low-lying eigenspectrum. Particularly noteworthy is
the potential for eigenenergies to merge as the strength of the imaginary
potential $\kappa $ is varied, leading to the emergence of the EP in the
spectrum. This intriguing behavior is illustrated in Figs. \ref{fig2}%
(a1-c1). Fig. \ref{fig2}(a1) distinctly depicts that when $\kappa =0$, the
reality of the low-lying energy spectrum remains intact. As $\kappa $\
surpasses $\kappa _{c_{1}}$, imaginary energies emerge from the two lowest
energies. When $\kappa $ exceeds $\kappa _{c_{2}}$, the real component
diminishes, causing these conjugate pairs to separate into four imaginary
eigenenergies.

\begin{figure*}[tbp]
\includegraphics[bb=63 20 1579 505,width=1.0\textwidth, clip]{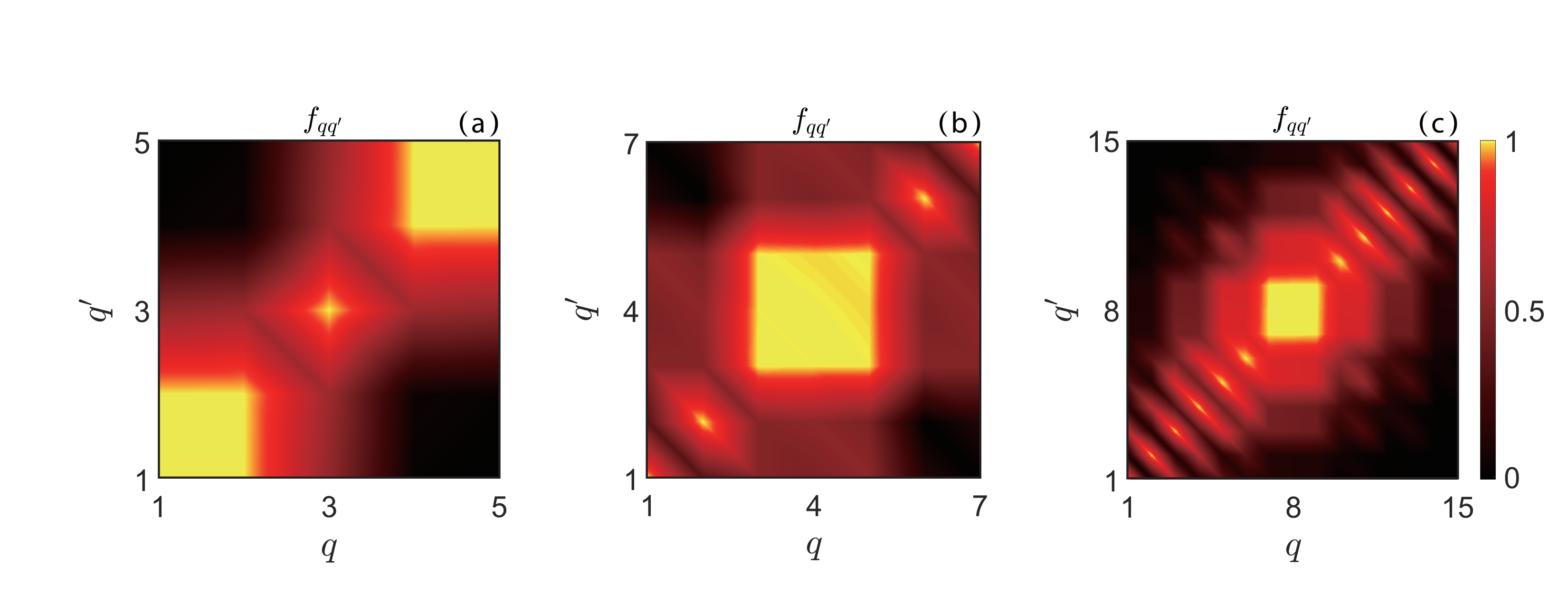}
\caption{(Color online) The overlap $f_{qq^{\prime }}$ of the eigenstates
within the low-lying spectrum of the discretized Hamiltonian $H$. The
strengths of the loss are (a) $\protect\kappa =0.0062$, (b) $\protect\kappa %
=0.0252$, and (c) $\protect\kappa =0.3440$, respectively. The other system
parameters within each pannel are the same as those in the systems labeled
by the green dashed lines in Figs. \protect\ref{fig2}(a2)-(c2),
respectively. }
\label{fig3}
\end{figure*}

To ascertain whether the eigenstates coalesce, we define the discretized
eigenstate as
\begin{equation}
\left\vert \psi _{q}\right\rangle =\underset{n=1}{\overset{N^{\prime }}{\sum
}}d_{q}\left( n\right) \left\vert x_{n}\right\rangle ,
\end{equation}%
and the fidelity between these states is given by
\begin{equation}
f_{qq^{\prime }}=\frac{1}{\Omega _{q}\Omega _{q^{\prime }}}\underset{n=1}{%
\overset{N^{\prime }}{\sum }}\left\vert d_{q}\left( n\right) \right\vert
\left\vert d_{q^{\prime }}\left( n\right) \right\vert ,
\end{equation}%
where $q$ and $q^{\prime }$ are indices that depict the ordering of the
imaginary components in the low-lying spectrum of the continuous system post
discretization. Here, $q$ $(q^{\prime })=1,2,\cdots ,2\omega +1$ and the
normalization factor is denoted as $\Omega _{q(q^{\prime })}=\sqrt{\underset{%
n}{\sum }\left\vert d_{q(q^{\prime })}\left( n\right) \right\vert ^{2}}$. In
Fig. \ref{fig3}, we numerically examine the overlap of the eigenstates in
the low-lying spectrum of the discretized Hamiltonian $H$. The system
parameters are consistent with those shown in Fig. \ref{fig2}. The yellow
shaded region indicates instances where $f_{qq^{\prime }}=1$. The
self-overlap along the diagonal represents a trivial region, while the
non-diagonal yellow region signifies a non-trivial region. Notably, an EP$2$
is observed in the case of $N=5$. Additionally, an EP$3$ is observed in the
cases of $N=7$ and $15$.

To streamline the ensuing discussion, let us first derive the effective
Hamiltonian concerning the low-lying spectrum. In the spirit of the
tight-binding approximation, the eigenstates within the low-lying spectrum
of $H$ and $H^{\dagger }$ can be given as
\begin{equation}
\left\vert \psi _{n}\right\rangle =\sum_{n^{\prime }=1}^{N^{\prime
}}a_{n,n^{\prime }}\left\vert x_{n^{\prime }}\right\rangle
=\sum_{l=1}^{N}d_{n,l}\left\vert \chi _{l}\right\rangle ,
\end{equation}%
and%
\begin{equation}
\left\vert \overline{\psi }_{n}\right\rangle =\sum_{n^{\prime
}=1}^{N^{\prime }}\overline{a}_{n,n^{\prime }}|x_{n^{\prime }}\rangle
=\sum_{l=1}^{N}\overline{d}_{n,l}|\chi _{l}\rangle ,
\end{equation}%
where $n=1,2,3,\cdots ,N$ complying with the biorthonormal relation $%
\left\langle \overline{\psi }_{m}|\psi _{n}\right\rangle =\delta _{mn}$. The
initial equality in the above expressions suggests that the low-lying
eigenstate can be expanded using a set of discretized states $\left\vert
x_{n^{\prime }}\right\rangle $ in the coordinate space. The subsequent
equality indicates that the low-lying eigenstate can be obtained by
combining the normalized ground state $\left\vert \chi _{l}\right\rangle $
of $l$-th equivalent single-well systems, essentially through a linear
combination of $N$ Wannier functions. By introducing the projection operator
\begin{equation}
P=\sum_{l=1}^{N}\left\vert \chi _{l}\right\rangle \langle \chi _{l}|,
\end{equation}%
and subsequently applying it to the Eq. (\ref{H0}). The effective
Hamiltonian in the $\{\left\vert \chi _{l}\right\rangle \}$ representation
can be given as
\begin{eqnarray}
h_{\text{\textrm{eff}},ll^{\prime }} &=&\sum_{m^{\prime },n^{\prime
}}^{N^{\prime }}\langle \chi _{l}|P^{-1}|x_{m^{\prime }}\rangle H_{m^{\prime
}n^{\prime }}^{0}\langle x_{n^{\prime }}|P|\chi _{l^{\prime }}\rangle  \notag
\\
&=&\sum_{n=1}^{N}E_{n}\langle \chi _{l}|\psi _{n}\rangle \langle \overline{%
\psi }_{n}|\chi _{l^{\prime }}\rangle .
\end{eqnarray}%
After some straightforward algebras, the effective tight-binding Hamiltonian
can be written as
\begin{equation}
H_{\text{\textrm{eff}}}=J_{\text{\textrm{eff}}}\sum_{j=1}^{N-1}(c_{j+1}^{%
\dagger }c_{j}+\text{\textrm{h.c.}})+iF_{\text{\textrm{eff}}}\sum_{j=1}^{N}[%
\frac{-(N+1)}{2}+j]c_{j}^{\dagger }c_{j},  \label{Heff}
\end{equation}%
where $c_{j}^{\dag }(c_{j})$ denotes\ the creation (annihilation) fermionic\
operator at the $j$-th site, and $N$ denotes the length of the system. The
effective hopping and tilted potential are given by
\begin{equation}
J_{\text{\textrm{eff}}}=\sum_{n=1}^{N}E_{n}\langle \chi _{l}|\psi
_{n}\rangle \langle \overline{\psi }_{n}|\chi _{l+1}\rangle
\end{equation}%
and%
\begin{equation}
F_{\text{\textrm{eff}}}=-i\sum_{n=1}^{N}\frac{E_{n}\langle \chi _{l}|\psi
_{n}\rangle \langle \overline{\psi }_{n}|\chi _{l}\rangle }{-(N+1)/2+l}
\end{equation}%
holds for any index value of $l=1,2,...,N.$

To simplify, we will omit the subscript of $J_{\text{\textrm{eff}}}$ and $F_{%
\text{\textrm{eff}}}$ in the following expressions. The Nambu expression of $%
H_{\text{\textrm{eff}}}$ can be represented as
\begin{equation}
H_{\text{\textrm{eff}}}=\varphi ^{T}h_{\text{\textrm{eff}}}\varphi ,
\end{equation}%
where
\begin{equation}
h_{\text{\textrm{eff}}}=\left(
\begin{array}{ccccc}
-2iF & J & 0 & 0 & 0 \\
J & -iF & J & 0 & 0 \\
0 & J & 0 & J & 0 \\
0 & 0 & J & iF & J \\
0 & 0 & 0 & J & 2iF%
\end{array}%
\right) ,  \label{heff}
\end{equation}%
and the base is represented as $\varphi =\left( c_{1}^{\dagger
},c_{2}^{\dagger },\cdots ,c_{5}^{\dagger }\right) ^{T}$. Considering the
secular equation%
\begin{equation}
h_{\text{\textrm{eff}}}\Psi _{\rho \sigma }=e_{\rho \sigma }\Psi _{\rho
\sigma },  \label{e_heff}
\end{equation}%
one can determine the eigenenergies of the system as%
\begin{equation}
e_{\rho \sigma }=\rho \lbrack (-5F^{2}+4J^{2}+\sigma \Delta )/2]^{1/2},
\end{equation}%
where $\Delta =[(2J^{2}-3F^{2})^{2}-12J^{2}F^{2}]^{1/2}$. Here $\Xi _{\rho
\sigma }=e_{\rho \sigma }+2iF$, with $\rho (\sigma )=\pm ,$ $0,$ and $%
e_{++}, $ $e_{+-},$ $e_{00},$ $e_{-+},$ $e_{--}$ are five eigenvalues of $h_{%
\text{\textrm{eff}}}$. The corresponding eigenstates are given by
\begin{equation}
\Psi _{\rho \sigma }=[1/(\Lambda _{\rho \sigma })^{1/2}](\Theta _{1},\Theta
_{2},\Theta _{3},\Theta _{4},\Theta _{5})^{T},
\end{equation}%
with the normalization coefficient being
\begin{align}
\Lambda _{\rho \sigma }& =J^{6}+|\Xi _{\rho \sigma }|^{2}J^{4}+|\Xi _{\rho
\sigma }\left( \Xi _{\rho \sigma }-iF\right) J-J^{3}|^{2}  \notag \\
& +\left( |\Xi _{\rho \sigma }|^{2}-3F^{2}\right) \left( |\Xi _{\rho \sigma
}|^{2}-2J^{2}\right) ^{2}[1+J^{2}\left( |\Xi _{\rho \sigma }|\right) ^{-2}],
\end{align}%
The other coefficients are listed as follow
\begin{eqnarray}
\Theta _{1} &=&J^{3},  \notag \\
\Theta _{2} &=&\Xi _{\rho \sigma }J^{2},  \notag \\
\Theta _{3} &=&\Xi _{\rho \sigma }\left( \Xi _{\rho \sigma }-iF\right)
J-J^{3},  \notag \\
\Theta _{4} &=&\left( \Xi _{\rho \sigma }-iF\right) \left( e_{\rho \sigma
}\Xi _{\rho \sigma }-2J^{2}\right) ,  \notag \\
\Theta _{5} &=&J\left( \Xi _{\rho \sigma }-iF\right) \left[ e_{\rho \sigma
}\Xi _{\rho \sigma }-2J^{2}\right] \left( \Xi _{\rho \sigma }-4iF\right)
^{-1}
\end{eqnarray}%
Notice that when the system parameters satisfy $\Delta =0$, we can obtain
the following relation
\begin{equation}
(F/J)_{c_{1}}=\pm \lbrack (6+3\sqrt{3})/2]^{-1/2}\approx \pm 0.423
\end{equation}%
and%
\begin{equation}
(F/J)_{c_{2}}=\pm \lbrack (6-3\sqrt{3})/2]^{-1/2}\approx \pm 1.577.
\end{equation}%
Evidently, two types of EP emerge in $h_{\text{\textrm{eff}}}$ when $%
(F/J)_{c_{1}}\approx \pm 0.423$ and $(F/J)_{c_{2}}\approx \pm 1.577$. These
are illustrated in Fig. \ref{fig2}(a2) and are consistent with those in the
low-lying spectrum of the discretized Hamiltonian $H$, and are further
verified by the coalescence of the eigenstates. We also present numerical
results for the cases of $N=7$ and $15$ in Figs. \ref{fig2} (b2)-(c2).
Additionally, three key features should be emphasized: (i) When $\left\vert
F/J\right\vert \leqslant (F/J)_{c_{1}}$, all the eigenvalues are real. (ii)
When $(F/J)_{c_{1}}<\left\vert F/J\right\vert \leqslant (F/J)_{c_{2}}$, the
eigenvalues become complex; (iii) When $\left\vert F/J\right\vert
>(F/J)_{c_{2}}$, the eigenvalues become imaginary, losing their real parts.
The above three regions correspond to the dynamics discussed in the
subsequent section.

So far, we determine the properties of EPs numerically and theoretically
through the $N=5$ non-Hermitian tight-binding model. Comparing Figs. \ref%
{fig2}(a1)-(c1) to Figs. \ref{fig2}(a2)-(c2), the low-lying spectrum of
discretized Hamiltonian $H$ shares the same structure with that of the
non-Hermitian effective Hamiltonian $H_{\text{\textrm{eff}}}$. The other
parameters are the same in Fig. \ref{fig2}.

\section{Scale-free EP}

\label{Scale-free EP} 
\begin{figure}[tbp]
\includegraphics[bb=0 0 864 935 ,width=0.5\textwidth, clip]{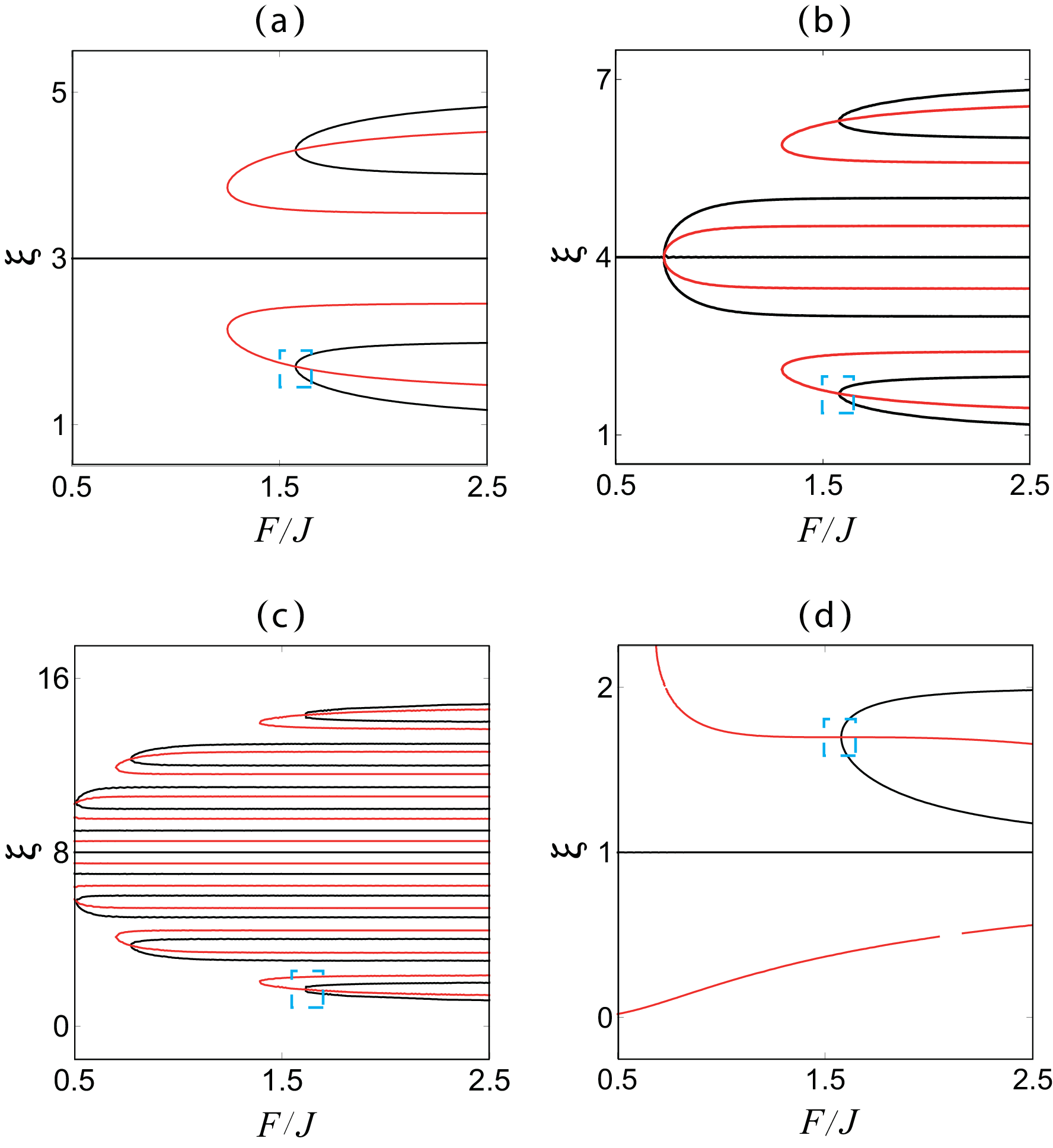}
\caption{(Color online) Plot of the solutions of $\protect\xi $ in Eqs. (%
\protect\ref{secular1}) and (\protect\ref{secular2}) as a function of $F/J$
based on the analytical solution. Figs. \protect\ref{fig4}(a)-(c) depict the
cases of $N=5$, $7$, and $15$, repectively. The black and red solid lines
represent the solutions of Eqs. (\protect\ref{secular1}) and (\protect\ref%
{secular2}), respectively. The intersection points within each figure
indicate the EP of the system. While $F/J$ decreases from infinity to a
finite value, the first EP, denoted by the blue dashed line, always emerges
at $(F/J)_{c_{2}}=1.577$. Fig. \protect\ref{fig4}(d) showcases the zero
solution of Eqs. (\protect\ref{r_secular1}) and (\protect\ref{hypergeometric}%
), confirming that the theoretical solution of Eq. (\protect\ref%
{hypergeometric}) in the limit of $\protect\alpha \rightarrow 0$, and $j-%
\protect\xi \rightarrow \infty $ is independent of the system size. This
independence suggests a scale-free EP at $(F/J)_{c_{2}}=1.577$ with $J$
assumed to be $1$ for simplicity.}
\label{fig4}
\end{figure}

In the previous section, it was demonstrated that the $%
H_{\text{\textrm{eff}}}$ effectively describes the low-energy behavior of
non-Hermitian continuous systems, particularly the EPs in the spectrum. In
this section, it will be shown that the system can host an EP independent of
the number of potential wells. We begin with the secular equation of the
effective Hamiltonian
\begin{equation}
h_{\text{\textrm{eff}}}|\Phi \rangle =e|\Phi \rangle ,  \label{s_heff}
\end{equation}%
where the eigenstate $|\Phi \rangle $ can be expressed as
\begin{equation}
|\Phi _{n}\rangle =\underset{j=1}{\overset{N}{\sum }}\beta _{j}\left\vert
j\right\rangle ,
\end{equation}%
with $\left\vert j\right\rangle =c_{j}^{\dagger }\left\vert \mathrm{Vac}%
\right\rangle $. Substituting the above equation into Eq. (\ref{s_heff}),
the following recurrence relation can be derived%
\begin{equation}
\beta _{j+1}+\beta _{j-1}=\frac{j-\xi }{\alpha }\beta _{j},
\end{equation}%
where $\alpha =iJ/F$, and $\xi =\varepsilon /F+(N+1)/2$ with $\varepsilon
=-ie$. Typically, the Bessel function satisfies the above recurrence
relation
\begin{equation}
Z_{n+1}\left( x\right) +Z_{n-1}\left( x\right) =\frac{2n}{x}Z_{n}\left(
x\right) ,
\end{equation}%
where $Z=\mathcal{J}$ $(Y)$ represents the first (or second) Bessel
function. It is important to note that the above equation remains valid for
purely imaginary quantities (the proof is provided in Appendix A). The
expansion coefficients $\beta _{j}$ will be expressed in terms of a
combination of Bessel functions as
\begin{equation}
\beta _{j}=A\mathcal{J}_{j-\xi }\left( 2\alpha \right) +BY_{j-\xi }\left(
2\alpha \right) ,
\end{equation}%
where the variables are transformed, i.e., $n\rightarrow j-\xi ,$ and $%
x\rightarrow 2\alpha $. The boundary condition of $\beta _{0}=\beta _{N+1}=0$
lead to the following equations%
\begin{equation}
A\mathcal{J}_{-\xi }\left( 2\alpha \right) +BY_{-\xi }\left( 2\alpha \right)
=0,
\end{equation}%
and
\begin{equation}
A\mathcal{J}_{N+1-\xi }\left( 2\alpha \right) +BY_{N+1-\xi }\left( 2\alpha
\right) =0.
\end{equation}%
By combining the above two equations, we obtain
\begin{equation}
\mathcal{J}_{-\xi }\left( 2\alpha \right) Y_{N+1-\xi }\left( 2\alpha \right)
-\mathcal{J}_{N+1-\xi }\left( 2\alpha \right) Y_{-\xi }\left( 2\alpha
\right) =0.  \label{secular1}
\end{equation}%
The order of the Bessel function $\xi $ must be real such that the
eigenvalues are imaginary numbers, which are determined by the zero points
of the above function. As $\alpha \rightarrow 0$, and $j-\xi \rightarrow
\infty $, we can obtain%
\begin{equation}
\mathcal{J}_{j-\xi }\left( -i\alpha \right) =\frac{1}{\sqrt{2\pi }}\left(
-ie\alpha /2\right) ^{j-\xi }\left( j-\xi \right) ^{-(j-\xi +\frac{1}{2})}
\end{equation}%
and%
\begin{equation}
\mathcal{J}_{j-\xi }\left( \alpha \right) =i^{(j-\xi )}\mathcal{J}_{j-\xi
}\left( -i\alpha \right) .
\end{equation}%
By utilizing the asymptotic behavior of the Bessel function, which is
detailed in Appendix B, we obtain
\begin{equation}
\mathcal{J}_{j-\xi }\left( 2\alpha \right) =\frac{1}{\sqrt{2\pi }}\left(
e\alpha \right) ^{j-\xi }\left( j-\xi \right) ^{-(j-\xi +\frac{1}{2})},
\end{equation}%
and
\begin{equation}
Y_{j-\xi }\left( 2\alpha \right) =\sqrt{\frac{2}{\pi }}\left( e\alpha
\right) ^{-\left( j-\xi \right) }\left( j-\xi \right) ^{j-\xi -\frac{1}{2}}.
\end{equation}%
This results in $\left\vert \mathcal{J}_{N+1-\xi }\left( 2\alpha \right)
\right\vert =0$ and $\left\vert Y_{N+1-\xi }\left( 2\alpha \right)
\right\vert \gg 1$. Therefore, the Eq. (\ref{secular1}) is simplified to
\begin{equation}
\mathcal{J}_{-\xi }\left( 2\alpha \right) =0.  \label{r_secular1}
\end{equation}%
By differentiating Eq. (\ref{secular1}) with respect to $\xi $, we have
\begin{equation}
\partial \left[ \mathcal{J}_{-\xi }\left( 2\alpha \right) Y_{N+1-\xi }\left(
2\alpha \right) -\mathcal{J}_{N+1-\xi }\left( 2\alpha \right) Y_{-\xi
}\left( 2\alpha \right) \right] /\partial \xi =0,  \label{secular2}
\end{equation}%
and combining the asymptotic behavior, we immediately obtain the additional
condition to determine the EP of the system
\begin{equation}
\partial \mathcal{J}_{-\xi }\left( 2\alpha \right) /\partial \xi =0.
\label{r_secular2}
\end{equation}%
Straightforward algebra shows that
\begin{eqnarray}
\partial \mathcal{J}_{-\xi }\left( 2\alpha \right) /\partial \xi &=&-\frac{%
\pi \alpha ^{-2\xi }}{2\Gamma ^{2}(-\xi +1)}[Y_{-\xi }\left( 2\alpha \right)
-\cot \left( -\xi \pi \right) \times  \notag \\
&&\mathcal{J}_{-\xi }\left( 2\alpha \right) ]_{2}F_{3}(-\xi ,-\xi +\frac{1}{2%
};-\xi +1,-\xi +  \notag \\
&&1,-2\xi +1;-4\alpha ^{2})-\mathcal{J}_{-\xi }\left( 2\alpha \right) \times
\lbrack -\frac{1}{2\xi }-  \notag \\
&&\psi \left( -\xi +1\right) +\ln \alpha +\alpha ^{2}/(\xi
^{2}-1)_{3}F_{4}(1,1,  \notag \\
&&\frac{3}{2};2,2,2+\xi ,-\xi +2;-4\alpha ^{2})],
\end{eqnarray}%
where $_{p}F_{q}$ represents the generalized hypergeometric function. By
neglecting the last three terms, we obtain
\begin{equation}
_{2}F_{3}(-\xi ,-\xi +\frac{1}{2};-\xi +1,-\xi +1,-2\xi +1;-4\alpha ^{2})=0.
\label{hypergeometric}
\end{equation}%
It is crucial to highlight that the determination of Eq. (\ref{r_secular1})
and Eq. (\ref{hypergeometric}) is independent of the system size $N$,
indicating a scale-free EP (the observation is presented in Fig. \ref{fig4}%
(d)). More details of the observations are further illustrated in Figs. \ref%
{fig4}(a)-(c), where the black and red solid lines correspond to the
solutions of Eqs. (\ref{secular1}) and (\ref{secular2}), respectively. The
intersection of these lines give rise to the EP of the system. The
intersection point, denoted by the blue dashed rectangle, consistently
occurs at $(F/J)_{c_{2}}=1.577$ for $N=5,$ $7$, and $15$ respectively. This
uniform behavior is in line with our theoretical forecasts, indicating a
universal trait.

\section{The dynamics at EP}

\label{The dynamics at EP} Now, we shift our focus to the dynamics of the
system. In the large $N$ limit, the system's dynamics remain unaffected by
the boundary condition due to the absence of the non-Hermitian skin effect.
This leads us to consider a system with periodic boundary conditions for
simplicity.

\begin{figure}[tbp]
\includegraphics[bb=27 44 1021 1036,width=0.5\textwidth, clip]{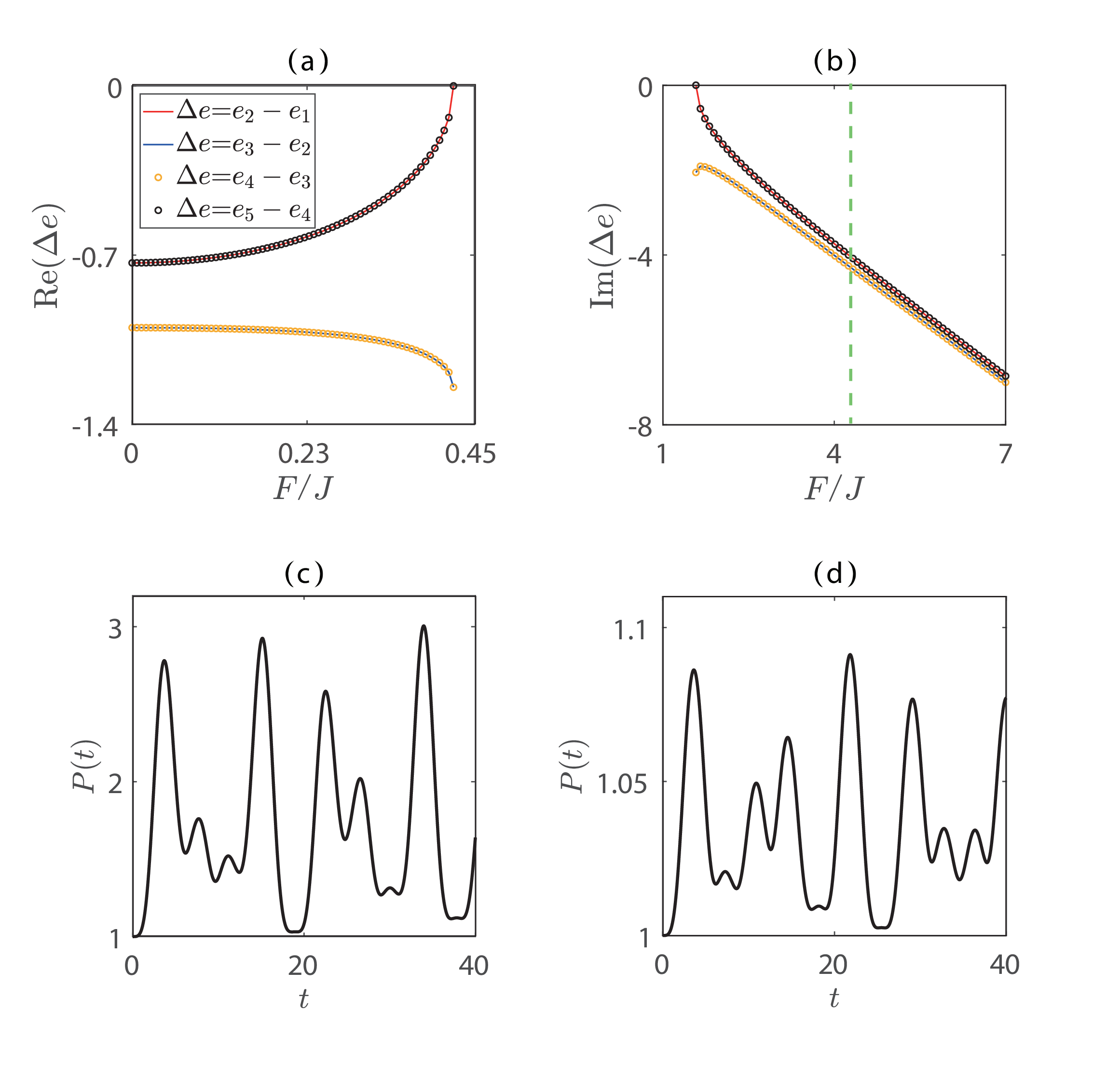}
\caption{(Color online) Energy spacing $\Delta e$ and the total Dirac
probability $P\left( t\right) $ of the effective Hamiltonian $H_{\mathrm{eff}%
}$. Figs. \protect\ref{fig5}(a)-(b) illustrate the full real and imaginary
energy spectra as $F/J$ varies. For $F/J>(F/J)_{c_{2}}$, the energy spectrum
displays equidistant spacing, while this pattern is absent in other cases.
The green dashed lines represent the situation when $F/J=4.3$. Figs. \protect
\ref{fig5}(c)-(d) display the numerical variations of $P(t)$ with respect to
$t$ for the systems with $F/J=0.2$, and $0.05$, respectively. The periodic
behavior diminishes as the equidistant energy spacing is not present.}
\label{fig5}
\end{figure}

\begin{figure}[tbp]
\includegraphics[bb=67 80 1096 1093,width=0.5\textwidth, clip]{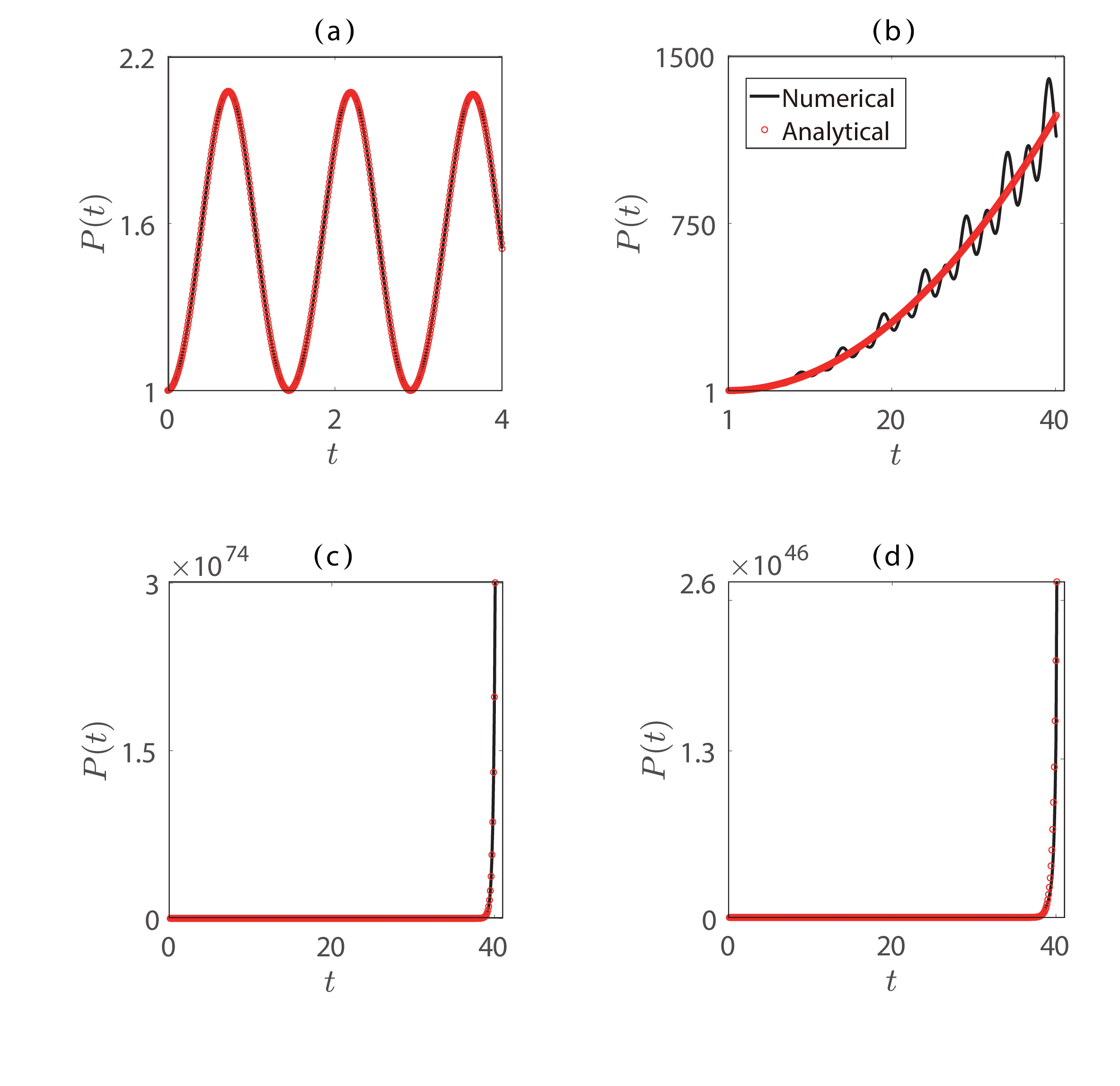}
\caption{(Color online) Plots illustrating of the Dirac probability $P(t)$
as a function of $t$. (a) For $F/J=4.3$ ($N=5$), representing the full real
spectrum, $P(t)$ displays oscillations with a period $T\approx 1.46$. (b) At
$(F/J)_{c_{1}}=0.423$ ($N=5$) and (c) $(F/J)_{c_{2}}=1.577$ ($N=5$),
indicating the first and second EPs of the systems respectively, $%
P(t)\varpropto t^{2}$ exemplifies a power law behavior. (d) With $F/J=0.732$
($N=7$). EP3 emerges as $P(t)\varpropto t^{4}$. The remaining system
parameters align with those of the green dashed line in Fig. \protect\ref%
{fig2}(b2). The red empty circle represents the analytical result, in good
agreement with the numerical simulation depicted by the solid black line.}
\label{fig6}
\end{figure}

Before delving into a detailed analysis of EP dynamics, we first explore the
system's dynamics significantly distant from the EP. It is imperative to
highlight that the derivation of the propagator relies on an equidistant and
complete real energy spectrum. In Figs. \ref{fig5}(a)-(b), we examine the
energy spacing $\Delta e$ of the Hamiltonian $H_{\mathrm{eff}}$ by adjusting
$F/J$. Two distinct regimes merit attention: (i) For $F/J<(F/J)_{c_{1}}$,
the system exhibits a complete real spectrum without the characteristic of
equidistant energy spacing. (ii) Conversely, for $F/J>(F/J)_{c_{1}}$, the
system may showcase equidistant energy spacing alongside a complete
imaginary energy spectrum. Upon considering the time evolution of the
Hamiltonian $-iH_{\mathrm{eff}}$, the presence of equidistant energy spacing
and a complete real spectrum is confined to a specific parameter range,
which will be the focal point of our subsequent discussion. By performing
the Fourier transformation%
\begin{equation}
c_{j}=\frac{1}{\sqrt{N}}\sum_{k}e^{ikj}c_{k},
\end{equation}%
we rewrite the Eq. (\ref{Heff}) in the form of
\begin{equation}
H_{\xi }\left( k\right) =-iH_{\mathrm{eff}}\left( k\right) ,
\end{equation}%
where
\begin{equation}
H_{\mathrm{eff}}\left( k\right) =\sum_{k}[2J\cos \left( k\right) -iF\frac{%
(N+1)}{2}-F\frac{\partial }{\partial k}]c_{k}^{\dagger }c_{k}.
\end{equation}%
Correspondingly, the eigenstate of $H_{\xi }\left( k\right) $ can be
expressed as%
\begin{equation}
\left\vert \psi \left( k\right) \right\rangle =\sum_{k}\psi \left( k\right)
\left\vert k\right\rangle ,
\end{equation}%
with $\left\vert k\right\rangle =c_{k}^{\dagger }\left\vert \mathrm{Vac}%
\right\rangle $. It fulfills the following secular equation
\begin{equation}
-i2J\cos \left( k\right) \psi \left( k\right) +iF\frac{\partial }{\partial k}%
\psi \left( k\right) =\xi \psi \left( k\right) ,
\end{equation}%
where $\xi =-ie+F(N+1)/2$, and it is equidistant number. By utilizing the
periodic boundary condition of the wavefunction $\psi \left( k+2\pi \right)
=\psi \left( k\right) $, we have
\begin{equation}
\left\vert \psi \left( k\right) \right\rangle =\frac{1}{\sqrt{N}}\underset{k}%
{\sum }\exp \{-i[i\frac{2J}{F}\sin (k)+\frac{\xi }{F}k]\}\left\vert
k\right\rangle .
\end{equation}%
In the coordinate representation, it can be rewritten as
\begin{equation}
\left\vert \psi \left( m\right) \right\rangle =\frac{1}{N}\sum_{j,k}\exp
\{-i[\left( m-j\right) k+\frac{2iJ}{F}\sin \left( k\right) ]\}\left\vert
j\right\rangle .
\end{equation}%
In the large $N$ limit, the summation in the above can be replaced with an
integral, yielding
\begin{equation}
\left\vert \psi \left( m\right) \right\rangle =\sum_{j}\mathcal{J}_{\left(
m-j\right) }(-\frac{2iJ}{F})\left\vert j\right\rangle .
\end{equation}%
Here, the Bessel Function of order $n$ is expressed as
\begin{equation}
\mathcal{J}_{n}\left( z\right) =\frac{1}{2\pi }\int_{-\pi }^{\pi }\exp
\left\{ -i\left[ n\theta -z\sin \theta \right] \right\} \mathrm{d}\theta .
\end{equation}%
The dynamics of the system is fully captured by the propagator. However, the
existence of the imaginary potential generally gives rise to an imaginary
spectrum, which brings an exponentially growing total probability. To omit
such divergence, we define $U\left( t\right) =\exp \left( -iH_{\xi }t\right)
$ with $H_{\varsigma }=-iH_{\mathrm{eff}}$ to observe the relative amplitude
in each lattice site. In the coordinate space, the matrix element can be
given as
\begin{equation}
U_{n^{\prime }n}(t)=\mathcal{J}_{n^{\prime }-n}[-\frac{4iJ}{F}\sin (-\frac{Ft%
}{2})]e^{i(n^{\prime }-n)\left( \pi -Ft\right) /2-in^{\prime }Ft}
\label{propa}
\end{equation}%
wherein the property of the Bessel function
\begin{equation}
\sum\limits_{k=-\infty }^{\infty }J_{k}(z)J_{k+p}(z)e^{ika}=J_{p}[2z\sin (%
\frac{a}{2})]e^{ip(\pi -a)/2},
\end{equation}%
is utilized. Evidently, it characterizes the periodic oscillation with
period $T=2\pi /F$. Fig. \ref{fig6}(a) showcases the time evolution of the
total Dirac probability $P\left( t\right) $ for the Hamiltonian $H_{\xi }$
with the system parameter $F/J=4.3$. When the system satisfies the condition
of equidistant energy spacing and a complete real spectrum, $P\left(
t\right) $ exhibits oscillations with an approximate period of $T=1.46$. The
red empty circle and black solid line denote the analytical formula Eq. (\ref%
{propa}) and numerical results, respectively. However, for the effective
Hamiltonian $H_{\mathrm{eff}}$, satisfying such conditions proves challenging,
regardless of the variations in the parameter $F/J$. Consequently, the
corresponding propagator fails to capture the system dynamics. In the region
of a complete real spectrum, Figs. \ref{fig5}(c)-(d) illustrate the system
dynamics for the system parameters $F/J=0.2$, and $0.05$, respectively. In
such region, the finite size effect precludes the emergence of a spectrum
with equidistant energy spacing, leading to the disappearance of periodic
behavior.

Next, we shift our focus to the EP dynamics of $H_{\mathrm{eff}}$. The
discussion is separated into three parts for clarity.

(i) When the system parameter satisfies the critical condition of $%
(F/J)_{c_{1}}=[(6+3\sqrt{3})/2]^{-1/2}\approx 0.423$, the first EP\ emerges
where two eigenstates coalesce. To elucidate the EP dynamics, we introduce
the generalized eigenspace based on the Eqs. (\ref{heff}) and (\ref{e_heff}%
). The matrix representation $h_{\mathrm{eff}}$ transforms to
\begin{equation}
h_{\mathrm{eff}}^{(B)}=S^{-1}h_{\mathrm{eff}}S=\left(
\begin{array}{ccccc}
e_{++} & 1 & 0 & 0 & 0 \\
0 & e_{+-} & 0 & 0 & 0 \\
0 & 0 & e_{00} & 0 & 0 \\
0 & 0 & 0 & e_{-+} & 1 \\
0 & 0 & 0 & 0 & e_{--}%
\end{array}%
\right) ,
\end{equation}%
where $S=(\varphi _{++},$ $\Phi _{+-},$ $\varphi _{00},$ $\varphi _{-+},$ $%
\Phi _{--})$ and the corresponding diagonal elements are $%
e_{++}=e_{+-}=1.246 $, and $e_{-+}=e_{--}=-1.246$. Here $\varphi _{++}$($%
\varphi _{-+}$) is the generalized eigenvector of the eigenvector $\Phi
_{+-} $($\Phi _{--}$) following the relations
\begin{equation}
(h_{\mathrm{eff}}-e_{++}I)\Phi _{+-}=\varphi _{++}
\end{equation}%
and
\begin{equation}
(h_{\mathrm{eff}}-e_{-+}I)\Phi _{--}=\varphi _{-+}.
\end{equation}%
It is crucial to transform a non-Hermitian matrix with EPs into its standard
Jordan block form. This transformation is essential because it allows for
the derivation of the time-dependent amplitude through the systematic
solution of decoupled differential equations. Straightforward algebraic
operations reveal that
\begin{widetext}
\begin{equation}
S=\left(
\begin{array}{ccccc}
0.041-0.329i & 0.494-0.058i & -0.436 & 0.041+0.329i & -0.494-0.058i \\
0.329-0.376i & 0.705+0.016i & -0.368i & -0.329-0.376i & 0.705-0.016i \\
0.529 & 0.707 & 0.591 & 0.529 & -0.707 \\
0.329+0.376i & 0.705-0.016i & 0.368i & -0.329+0.376i & 0.705+0.016i \\
0.041+0.329i & 0.494+0.058i & -0.436 & 0.041-0.329i & -0.494+0.058i%
\end{array}%
\right) .
\end{equation}
\end{widetext}
Note that the transformation matrix $S$ is not unique and depends on the
choice of the generalized eigenvector. Considering the time evolution of the
arbitrary initial state $\psi \left( 0\right) =\left[ c_{1}\left( 0\right)
\text{, }c_{2}\left( 0\right) \text{, }c_{3}\left( 0\right) \text{, }%
c_{4}\left( 0\right) \text{, }c_{5}\left( 0\right) \right] ^{T}$, the
evolved state is governed by $h_{\mathrm{eff}}^{(B)}$, i.e., $i\partial
_{t}\psi =h_{\mathrm{eff}}^{(B)}\psi $. The amplitude of the evolved state
satisfies the following system of differential equations%
\begin{equation}
\left\{
\begin{array}{c}
i\partial _{t}c_{1}(t)=e_{++}c_{1}(t)+c_{2}(t) \\
i\partial _{t}c_{2}(t)=e_{+-}c_{2}(t) \\
i\partial _{t}c_{3}(t)=0 \\
i\partial _{t}c_{4}(t)=e_{--}c_{4}(t)+c_{5}(t) \\
i\partial _{t}c_{5}(t)=e_{-+}c_{5}(t)%
\end{array}%
\right. ,
\end{equation}%
leading to the solution
\begin{equation}
\psi \left( t\right) =\left(
\begin{array}{c}
c_{1}(0)e^{-1.246it}-ic_{2}(0)te^{-1.246it} \\
c_{2}\left( 0\right) e^{-1.246it} \\
c_{3}\left( 0\right) \\
c_{4}(0)e^{1.246it}-ic_{5}(0)te^{1.246it} \\
c_{5}\left( 0\right) e^{1.246it}%
\end{array}%
\right) .
\end{equation}%
Taking $c_{j}(0)=\delta _{j,3}$, and transforming it to the coordinate space
$S^{-1}|\psi \left( t\right) \rangle $ results in
\begin{equation}
\psi \left( t\right) =\left(
\begin{array}{c}
-0.606e^{-1.246it}+0.620ite^{-1.246it} \\
0.620e^{-1.246it} \\
1.293 \\
-0.606e^{1.246it}+0.620ite^{1.246it} \\
-0.620e^{1.246it}%
\end{array}%
\right) .
\end{equation}%
Evidently, the Dirac probability $P(t)=3.175+0.769t^{2}$ increases with time
$t$ according to a power law. It is the evidence of the EP2, which is also
demonstrated in Fig. \ref{fig6}(b).

(ii) When the system stays at the second EP, i.e., $(F/J)_{c_{2}}=[(6-3\sqrt{%
3})/2]^{-1/2}\approx 1.577$. The matrix form $h_{\mathrm{eff}}$ is given by
\begin{equation}
h_{\mathrm{eff}}=\left(
\begin{array}{ccccc}
-3.155i & 1 & 0 & 0 & 0 \\
1 & -1.577i & 1 & 0 & 0 \\
0 & 1 & 0 & 1 & 0 \\
0 & 0 & 1 & 1.577i & 1 \\
0 & 0 & 0 & 1 & 3.155i%
\end{array}%
\right) ,
\end{equation}%
Following the same procedures, we have
\begin{equation}
h_{\mathrm{eff}}^{(B)}=S^{-1}h_{0}S=\left(
\begin{array}{ccccc}
-2.054i & 1 & 0 & 0 & 0 \\
0 & -2.054i & 0 & 0 & 0 \\
0 & 0 & 0 & 0 & 0 \\
0 & 0 & 0 & 2.054i & 1 \\
0 & 0 & 0 & 0 & 2.054i%
\end{array}%
\right) ,
\end{equation}%
where
\begin{equation}
S=\left(
\begin{array}{ccccc}
-0.639i & -0.117 & -0.132 & 0.015i & -0.029 \\
0.703 & -0.768i & -0.416i & -0.079 & -0.139i \\
0.303i & 0.454 & 0.787 & -0.303i & 0.454 \\
-0.079 & 0.139i & 0.416i & 0.703 & 0.768i \\
-0.015i & -0.029 & -0.132 & 0.639i & -0.117%
\end{array}%
\right) .
\end{equation}%
For an arbitrary initial state, its evolved state can be given as
\begin{equation}
\psi \left( t\right) =\left(
\begin{array}{c}
c_{1}\left( 0\right) e^{-2.0543t}-ic_{2}\left( 0\right) te^{-2.0543t} \\
c_{2}\left( 0\right) e^{-2.0543t} \\
c_{3}\left( 0\right) \\
c_{4}\left( 0\right) e^{2.0543t}-ic_{5}\left( 0\right) te^{2.0543t} \\
c_{5}\left( 0\right) e^{2.0543t}%
\end{array}%
\right) .
\end{equation}%
Taking $c_{j}(0)=\delta _{j,3}$, and transforming it to the coordinate space
$S^{-1}|\psi \left( t\right) \rangle $ results in the Dirac probability $%
P(t)=[1.612\cosh (4.109t)]t^{2}-[1.128\sinh (4.109t)]t+1.810\cosh
(4.109t)+6.497$, increasing with time $t$ following a power law. Moreover,
the analysis indicates that $(F/J)_{c_{2}}\approx 1.577$ represents the EP2.
This phenomenon is further illustrated in Fig. \ref{fig5}(c).

(iii) Now we consider a high-order EP in a larger system. When $N=7$, the
system parameter $F/J\approx 0.732$ corresponding to the green dashed line
in Fig. \ref{fig2}(b2) admits the following core matrix
\begin{widetext}
\begin{equation}
h_{\mathrm{eff}}=\left(
\begin{array}{ccccccc}
-2.196i & 1 & 0 & 0 & 0 & 0 & 0 \\
1 & -1.464i & 1 & 0 & 0 & 0 & 0 \\
0 & 1 & -0.732i & 1 & 0 & 0 & 0 \\
0 & 0 & 1 & 0 & 1 & 0 & 0 \\
0 & 0 & 0 & 1 & 0.732i & 1 & 0 \\
0 & 0 & 0 & 0 & 1 & 1.464i & 1 \\
0 & 0 & 0 & 0 & 0 & 1 & 2.196i%
\end{array}%
\right) .
\end{equation}
\end{widetext}The eigenvalues are $e_{1}=-A-iB$, $e_{2}=A-iB$, $e_{3}=A+iB$,
$e_{4}=e_{5}=e_{6}=0$, and $e_{7}=-A+iB$ where $A=1.019,$ $B=1.337.$
Obviously, there are three coalescent eigenstates in the above equation.
Similarly, the standard Jordan Block form is written as%
\begin{equation}
h_{\mathrm{eff}}^{(B)}=S^{-1}h_{\mathrm{eff}}S=\left(
\begin{array}{ccccccc}
e_{1} & 0 & 0 & 0 & 0 & 0 & 0 \\
0 & e_{2} & 0 & 0 & 0 & 0 & 0 \\
0 & 0 & e_{3} & 0 & 0 & 0 & 0 \\
0 & 0 & 0 & e_{4} & 1 & 0 & 0 \\
0 & 0 & 0 & 0 & e_{5} & 1 & 0 \\
0 & 0 & 0 & 0 & 0 & e_{6} & 0 \\
0 & 0 & 0 & 0 & 0 & 0 & e_{7}%
\end{array}%
\right) ,
\end{equation}%
where
\begin{widetext}
\begin{equation}
S=\left(
\begin{array}{ccccccc}
-0.377-0.318i & 0.377-0.318i & 0.008+0.004i & 0.116i & -0.173 & -0.128i &
-0.008+0.004i \\
0.657 & 0.657 & -0.007+0.031i & -0.254 & -0.264i & 0.109 & -0.007-0.031i \\
-0.293+0.401i & 0.293+0.401i & -0.102+0.009i & -0.487i & 0.305 & 0.024i &
0.102+0.009i \\
-0.116-0.232i & -0.116+0.232i & -0.116-0.232i & 0.610 & 0 & 0.178 &
-0.116+0.232i \\
0.102-0.009i & -0.102-0.009i & 0.293-0.401i & 0.487i & 0.305 & -0.024i &
-0.293-0.401i \\
-0.007+0.031i & -0.007-0.031i & 0.657 & -0.254 & 0.264i & 0.109 & 0.657 \\
-0.008-0.004i & 0.008-0.004i & 0.377+0.318i & -0.116i & -0.173 & 0.128i &
-0.377+0.318i%
\end{array}%
\right) .
\end{equation}
\end{widetext}Thus the evolved state can be given as
\begin{equation}
\psi \left( t\right) =\left(
\begin{array}{c}
c_{1}(0)e^{-ie_{1}t} \\
c_{2}(0)e^{-ie_{2}t} \\
c_{3}(0)e^{-ie_{3}t} \\
-c_{6}(0)\frac{t^{2}}{2}-ic_{5}(0)t+c_{4}(0) \\
-ic_{6}(0)t+c_{5}(0) \\
c_{6}(0) \\
c_{7}(0)e^{-ie_{7}t}%
\end{array}%
\right) .
\end{equation}%
Taking $c_{j}(0)=\delta _{j,4}$, and transforming it to the coordinate space
$S^{-1}|\psi \left( t\right) \rangle $ results in the Dirac probability $%
P(t)=8.162t^{4}+29.627t^{2}+1.940\cosh (2.674t)+32.916$, increasing with time
$t$ following a power law. However, it scales proportionally to $t^{4}$
indicating the emergence of EP3. The numerical simulation is performed in
Fig. \ref{fig6}(d).

\section{Summary}

\label{Summary} In summary, we have systematically investigated the
non-Hermitian continuous model, wherein non-Hermiticity arises from an
imaginary on-site potential achievable through artificial synthetic systems.
Within the low-energy sector, the spectrum of the system can transition from
real to complex values. At the transition point, we have observed the
presence of EPs, and their characteristics can be effectively modeled using
a tilted imaginary potential within an adapted tight-binding Hamiltonian
framework. Furthermore, the effective Hamiltonian can exhibit a scale-free
EP, characterized by typical dynamic behavior. The growth of the total Dirac
probability over time conforms to a power law, with its exponent determined
by the order of the EP. These findings provide valuable insights into the
dynamics of EPs within practical experiments and pave the way for exploring
novel critical dynamic phenomena.

\section*{Appendix}

\subsection{Recurrence relation}

\label{A}

Given the Bessel function defined by the differential equation
\begin{equation}
x^{2}\frac{d^{2}R}{dx^{2}}+x\frac{dR}{dx}+(x^{2}-n^{2})R=0,
\end{equation}%
where $n$ represents the order of the function. $\mathcal{J}_{n}\left(
x\right) $ emerges as the solution to the $n$th-order Bessel function.

The series expression of the first kind Bessel function, denoted as $%
\mathcal{J}_{n}\left( x\right) $, is given by
\begin{equation}
\mathcal{J}_{n}\left( x\right) =\underset{k=0}{\overset{\infty }{\sum }}%
\frac{\left( -1\right) ^{k}}{k!}\frac{1}{\Gamma \left( n+k+1\right) }(\frac{x%
}{2})^{2k+n},
\end{equation}%
Multiplying both sides of the above equation by $x^{n},$ and taking the
first derivative with respect to $x$, we have
\begin{equation}
\frac{d}{dx}\left[ x^{n}\mathcal{J}_{n}\left( x\right) \right] =nx^{n-1}%
\mathcal{J}_{n}\left( x\right) +x^{n}\mathcal{J}_{n}^{^{\prime }}\left(
x\right) =x^{n}\mathcal{J}_{n-1}\left( x\right) ,  \label{AP1-3}
\end{equation}%
Similarly, we can obtain the relationship
\begin{equation}
\frac{d}{dx}\left[ x^{-n}\mathcal{J}_{n}\left( x\right) \right] =-x^{n}%
\mathcal{J}_{n+1}\left( x\right) .  \label{AP1-4}
\end{equation}%
By combining Eq. (\ref{AP1-3}) and Eq. (\ref{AP1-4}), and eliminating $%
\mathcal{J}_{n}^{^{\prime }}\left( x\right) $, we get
\begin{equation}
\mathcal{J}_{n+1}\left( x\right) +\mathcal{J}_{n-1}\left( x\right) =\frac{2n%
}{x}\mathcal{J}_{n}\left( x\right) ,
\end{equation}%
This recursive relation for $\mathcal{J}_{n}\left( x\right) $ holds true for
any $n$.

Considering the second kind Bessel function,
\begin{equation}
Y_{n}\left( x\right) =\frac{\cos n\pi \mathcal{J}_{n}\left( x\right) -%
\mathcal{J}_{-n}\left( x\right) }{\sin n\pi },
\end{equation}%
By substituting the series expression of $\mathcal{J}_{n}\left( x\right) $
into the above equation, we can show that $Y_{n}\left( x\right) $ also
satisfies a recursive relation. Letting $\xi =ix$, we can rewrite the Bessel
equation as
\begin{equation}
\xi ^{2}\frac{d^{2}R}{dx^{2}}+\xi \frac{dR}{dx}-\left( \xi ^{2}+n^{2}\right)
R=0,
\end{equation}%
where $\mathcal{J}_{n}\left( \xi \right) =\mathcal{J}_{n}\left( ix\right) $,
and its series expansion can be expressed as
\begin{equation}
\mathcal{J}_{n}\left( ix\right) =i^{n}\underset{k=0}{\overset{\infty }{\sum }%
}\frac{\left( -1\right) ^{k}}{k!}\frac{1}{\Gamma \left( n+k+1\right) }(\frac{%
x}{2})^{2k+n}.
\end{equation}%
The above expression indicates that the imaginary argument Bessel function
\begin{equation}
\mathcal{J}_{n}\left( ix\right) =i^{n}\mathcal{J}_{n}\left( x\right) ,
\end{equation}%
also satisfies the recursive relation of the real argument Bessel function.

\subsection{Asymptotic behavior}

\label{B}

Considering the series expression of the first kind Bessel function,
\begin{equation}
\mathcal{J}_{n}\left( x\right) =\underset{k=0}{\overset{\infty }{\sum }}%
\frac{\left( -1\right) ^{k}}{k!}\frac{1}{\Gamma \left( n+k+1\right) }(\frac{x%
}{2})^{2k+n}.
\end{equation}%
When $x\rightarrow 0$, the dominant term is given by%
\begin{equation}
\mathcal{J}_{n}\left( x\right) =\frac{1}{\Gamma \left( n+1\right) }(\frac{x}{%
2})^{n},
\end{equation}%
where
\begin{equation}
\Gamma \left( n+1\right) =n^{n}e^{-n}(2\pi n)^{1/2}[1+\frac{1}{12n}+o\left(
n^{-2}\right) ].
\end{equation}%
As $n\rightarrow \infty $, the expression simplifies to
\begin{equation}
\Gamma \left( n+1\right) =n^{n}e^{-n}(2\pi n)^{1/2}.
\end{equation}%
Therefore, the first kind Bessel function satisfies the relation
\begin{equation}
\mathcal{J}_{n}\left( x\right) =(2\pi n)^{-1/2}(\frac{ex}{2n})^{n},
\end{equation}%
where the parameters satisfy $x\rightarrow 0,$ $n\rightarrow \infty $.

Similarly, the second kind Bessel function satisfies the relation
\begin{equation}
Y_{n}\left( x\right) =-\frac{\Gamma \left( n\right) }{\pi }(\frac{x}{2}%
)^{-n},
\end{equation}%
Considering the recursive relation
\begin{equation}
n\Gamma \left( n\right) =\Gamma \left( n+1\right) ,
\end{equation}%
we can rewrite the second kind Bessel function as
\begin{equation}
Y_{n}\left( x\right) =(\frac{2}{\pi n})^{1/2}(\frac{ex}{2n})^{-n}.
\end{equation}

\acknowledgments We acknowledge the support of the National Natural Science
Foundation of China (Grants No. 12305026, No. 12275193).

\end{document}